\documentstyle[aaspp4,12pt,epsfig,color]{article}

\begin{document}

\title{Models of Disk Evolution: Confrontation with Observations}

\author{Rychard Bouwens}
\affil{Physics Department,
   University of California,
    Berkeley, CA 94720; bouwens@astro.berkeley.edu}
\centerline \&
\author {Joseph Silk}
\affil{Astrophysics, Department of Physics, University of Oxford, Oxford OX1 3RH UK and Astronomy and Physics Departments, and Center for Particle
    Astrophysics, University of California,
     Berkeley, CA 94720; silk@astro.berkeley.edu}

\begin{abstract}

We present simple models for disk evolution based on two different
approaches: a forward approach based on predictions generic to
hierarchical models for structure formation (e.g., Mo, Mao, \& White
1998) and a backwards approach based on detailed modeling of the
Milky Way galaxy (e.g., Bouwens, Cay\'on, \& Silk 1997).  We normalize
these models to local observations and predict high-redshift
luminosities, sizes, circular velocities, and surface brightnesses.
Both approaches yield somewhat similar predictions for size, surface
brightness, and luminosity evolution though they clearly differ in the
amount of number evolution.  These predictions seem to be broadly
consistent with the high-redshift observations of Simard et al.\
(1999), suggesting that the $B$-band surface brightness of disks has
indeed evolved by $\sim1.5^m$ from $z\sim0$ to $z\sim1$ similar to the
models and is not an artifact of selection effects as previously
claimed.  We also find a lack of low surface brightness galaxies in
several high redshift samples relative to model predictions based on
local samples (de Jong \& van der Kruit 1994; Mathewson, Ford, \&
Buchhorn 1992).

\end{abstract}


\section{Introduction}

Over the last several years, there has been a steady increase in the
number and quality of observations available for disk galaxies from
$z=0$ and $z=1$.  Schade et al.\ (1995,1996), using early ground and
space based images of galaxies from the Canada-France Redshift Survey
(CFRS), found a net increase in the surface brightness of galaxies to
$z\sim1$.  Along the same lines, Roche et al.\ (1998), compiling 347
galaxies from the Medium Deep Survey and other surveys, concluded that
disk galaxies had undergone a net evolution in surface brightness and
a net devolution in size.  Lilly et al.\ (1998), using structural
parameters extracted from HST images of the combined CFRS and LDSS2
sample, concluded that there has been essentially no evolution in
large disks out to $z\sim1$.  As a preliminary effort as part of the
DEEP survey, Vogt et al.\ (1996,1997) found little evolution in the
Tully-Fisher relationship ($<0.3^m$) out to $z\sim1$.  More recently,
these observations have been augmented by the DEEP sample with 197
galaxies from the Groth strip to $I<23.5$, 1.5 magnitudes deeper than
the LDSS2-CFRS sample.  In a first paper, Simard et al.\ (1998)
concluded that there had been little evolution in the disk surface
brightness distribution to $z\sim1$ contrary to previous claims.

A number of different approaches have been proposed for making
specific predictions about disk evolution.  Mo, Mao, \& White (1998a)
showed how the standard paradigm for hierarchical growth of structure
combined with simple assumptions about angular momentum conservation
led to simple scaling relationships for the change in disk properties
as a function of redshift.  Other authors (Ferrini et al.\ 1994;
Prantzos \& Aubert 1995; Prantzos \& Silk 1998; Boissier \& Prantzos
1999; Chiappini, Matteucci \& Gratton 1997), taking more of a
backwards approach to the problem, used detailed studies of the
profiles of the Milky Way and other nearby galaxies to propose
radially dependent models of star formation in disk galaxies, models
which could be used to make detailed predictions about high-redshift
disk evolution.

Already there have been a number of elegant studies in which both the
backwards approach (Cay\'on, Silk, \& Charlot 1996; Bouwens, Cay\'on,
\& Silk 1997; Roche et al.\ 1998) and the forwards approach (Mao, Mo,
\& White 1998; Steinmetz \& Navarro 1999; Contardo, Steinmetz, \&
Fritze-von Alvensleben 1998; van den Bosch 1998; Mo, Mao, \& White
1998b) have been used to interpret the observations available for disk
galaxies, mostly to $z\sim1$.

Unfortunately, none of these studies considered the important effect
that a large spread in surface brightness could have on the
interpretation of these observations, particularly the potentially
large fraction of low surface brightness galaxies.  In some studies,
the surface brightness selection effects at low and high redshift were
simply ignored, and in others, e.g., Roche et al.\ (1998), the spread
was limited to $0.3\,\textrm{mag/arcsec}^2$ about Freeman's law
(Freeman 1970).  Clearly, given the apparent large numbers of low
surface brightness galaxies seen locally, it is quite logical to
wonder if these galaxies are detectable in current high redshift
surveys.  Indeed, one might wonder whether these galaxies or the
observed correlation between luminosity and surface brightness may
have already affected the interpretation of high redshift
observations.  In light of the recent claim by Simard et al.\ (1999)
that the apparent surface brightness evolution thus far inferred to
$z\sim1$ is completely due to surface brightness selection effects,
such a study would seem to be especially timely.  Secondly, none of
these studies directly compared the predictions of the forward and
backward approaches using the same observations.  Simple comparisons
of the scaling expected in surface brightness, size, luminosity, and
number are useful for interpreting the high redshift observations.

To address these shortcomings, we shall therefore consider
implementations of both approaches, normalize them to the observed
$z\sim0$ size-luminosity relationship, compare their predictions, and
consider how each of them fares at explaining the observed disk
evolution out to $z\sim1$ incorporating all the selection effects as
they are best understood.  We commence by presenting our models (\S2)
and the observational samples with which we compare (\S3).  We present
the results (\S4), discuss them (\S5), and then summarize our
conclusions (\S6).  Throughout this study, we use $H_0 =
50\,\textrm{km/s/Mpc}$ unless otherwise noted.

\section{Models}

We begin by sketching the base $z=0$ model to be used for both the
models which follow.  We use a set of gaussian LFs based on those
presented in Binggeli, Sandage, and Tammann (1988):
\begin{equation}
\phi(M) dM = \frac{\phi_0}{2\pi \sigma_M}
\exp(-(\frac{M-M_{*}}{\sigma_M})^2) dM
\end{equation}
We adjusted the bulge-to-total ($B/T$) distributions of these galaxy
types to obtain fair agreement with the de Jong \& van der Kruit
(1994) sample.  Finally, we adjusted the luminosity function so there
was rough agreement with the combined Sabc and Sdm luminosity
functions presented in Pozzetti, Bruzual, \& Zamorani (1996).  We
present our parameterized populations in Table 1.  

\begin{deluxetable}{cccccccc}
\tablewidth{0pt}
\tablecaption{Model parameters for disk population.\label{lowl}}
\tablehead{
\colhead{$\phi_o$} & \colhead{$\sigma_M$} & \colhead{$M_{b_j}^*$} &
\colhead{$B/T$} &
\colhead{$\sigma_{B/T}$}}
\startdata
2.0 & 1.1 & -19.4 & 0.25 & 1.8\\
2.5 & 1.1 & -19.6 & 0.08 & 0.5\\
8.0 & 1.3 & -18.4 & 0.04 & 0.25\\
24.0 & 1.3 & -16.0 & 0.01 & 0.25
\enddata
\end{deluxetable}

For the above luminosity functions, we convert $z=0$ $B$-band
luminosity to mass using a constant mass-to-light ratio, where the
mass of a $M_{b_J} = -21.1$ galaxy is $1.1 \cdot 10^{12} M_{\odot}$.
We assume a log-normal scatter of 0.3 dex to reproduce the observed
Tully-Fisher scatter though variation in the formation times (van den
Bosch 1998) and concentration indexes (Avila-Reese, Firmani, \&
Hernandez 1998) certainly play a role.

To translate this mass into a circular velocity and size, we calculate
the time at which the ambient halo formed.  Since halos are always
accreting more mass and merging with larger halos, there is some
ambiguity in defining this, so for simplicity we take it to equal the
redshift at which half of the mass in a halo has been assembled.  We
determine the distribution of formation times using the procedure
outlined in Section 2.5.2 of Lacey \& Cole (1993).  We take the
circular velocity of the halo to be that corresponding to the halo at
its formation time using Eq. (14).

Then, given the mass, circular velocity, and luminosity of the $z=0$
disk, we randomly draw the sizes $r_e$ from the following
distribution:
\begin{equation}
\phi (r_e) d \log r_e = \frac{1}{\sigma_{\lambda}\sqrt{2\pi}} 
\exp(-\frac{1}{2} [\frac{\log r_e/r_e ^{*} - 0.4(M-M_{*,s})(1/3)}
{\sigma_{\lambda}/\ln(10)}]^2) d \log r_e
\end{equation}
where $r_e^* = 6.9 \textrm{kpc}$, $\sigma_{\lambda} = 0.37$, and $M_{*,s}
= -21.1$.  We provide a basic observational and theoretical motivation
for this scaling in \S2.1.  Note that here the surface brightness is
proportional to $L^{1/3}$, that the spread in the size distribution is
proportional to the spread in the distribution of $\lambda$, and that
the scale length (surface brightness) of the average $L_{*}$ galaxy is
exactly equal to that predicted by Freeman's law.

We assume that the SED of disks and bulges is identical to that of a
10 Gyr-old stellar population with an e-folding time of 4.5 Gyr and a
$\tau_{B} = 0.3$ foreground dust screen, the extinction curve being
that of Calzetti (1997).  For simplicity, we assume this SED is
constant independent of time.  We use the Bruzual \& Charlot tables
from the Leitherer et al.\ (1996) compilation for this calculation.
In the following models, we evolve the size, number, and luminosity of
all galaxy types using simple single-valued functions of redshift:
\begin{equation}
R(z) = R(0) E_R (z)
\end{equation}
\begin{equation}
N(z) = N(0) E_N (z)
\end{equation}
\begin{equation}
L(z) = L(0) E_L (z)
\end{equation}
For simplicity, we assume similar scalings in the properties of bulges
as a function of time.

Since the color and luminosity of disks has been shown to be
correlated with inclination, it is reasonable to suppose that disks
are not transparent.  Unfortunately, there is much controversy
concerning the degree to which disks are or are not transparent.  For
better or worse, we will side-step this controversy and simply adopt
the Tully \& Fouqu\'e (1985) prescription for extinction in the $B$
band:
\begin{equation}
A_B = -2.5 \log(f(1-\exp(-\tau \sec i)) + (1 - 2f)
(\frac{1-\exp(-\tau \sec i)}{\tau \sec i}))
\end{equation}
where $\tau = 0.55$, $f = 0.25$, and $i$ is the inclination of the
disk, 0 corresponding to a face-on disk.  We shall assume our model
galaxies are always observed at an inclination of $70\deg$ and
therefore always correct the observations to this inclination for
comparison with the models.  In the $B$ band, this corresponds to an
extinction correction of $0.67^m$.

\subsection{Hierarchical Model (Forwards Approach)}

The use of simple scaling relationships between the properties of
disks and the halos in which they live has provided a relatively
successful way of explaining both the internal correlations between
disk properties and their evolution to high redshift.  In this picture
developed by Fall \& Estathiou (1980) and revived more recently by
Dalcanton et al.\ (1997) and Mo et al.\ (1998) among others, the
bivariate mass and angular momentum distribution nicely translates
into a luminosity and surface brightness relationship for disk
galaxies, mass translating directly into luminosity and the
dimensionless angular momentum translating directly into surface
brightness.

This picture has had much success in explaining the internal
correlations between size, circular velocity, and mass.  For example,
De Jong \& Lacey (1999) recently showed that the observed local
bivariate luminosity-size distribution is nicely fit by this picture,
albeit with a slightly smaller scatter in surface brightness than
might otherwise be expected.  For constant mass-to-light ratios, the
rough $M \propto V_c^3$ relationship for halos provides a relatively
natural explanation for the luminosity-circular velocity
(Tully-Fisher) relationship (Dalcanton, Spergel, \& Summers 1997a; Mo
et al.\ 1998a; Steinmetz \& Navarro 1999; Contardo et al.\ 1998).
Finally, the $R \propto V_c$ relationship found in galaxy halos is
similarly observed in the disk population (Courteau 1997).

On the other hand, this picture says little, if anything, about how
the gas disk evolved over time and therefore what its local properties
(i.e., spatial variations in the metallicity, color, stellar ages, gas
density, etc.) or global properties (total gas mass) are, and
therefore comparisons of this sort will depend upon the model adopted,
whether it be one of the popular semi-analytic approaches (Cole et
al.\ 1999, Somerville \& Primack 1998) or a full N-body hydrodynamical
simulation (e.g., Contardo et al.\ 1998).  Moreover, it is now
apparent that the simple scaling model is fundamentally flawed with
regard to the implementation of galaxy formation theory as revealed by
high resolution numerical simulations (Moore et al. 1999; Navarro \&
Steinmetz 1999; Steinmetz \& Navarro 1999).  Since it however is the
only detailed model available, it is imperative to fully explore
comparisons with data, properly incorporating observational selection
effects, in order to establish the correct basis for ultimately
refining the model.

For a detailed discussion of this picture, i.e., the idea that simple
scalings in the properties of halos lead to simple scalings in the
properties of disks, the reader is referred to Mo et al.\ (1998a) and
later papers by the same authors.  For the sake of clarity, we shall
review some of this material.  In the standard spherical collapse
model for an Einstein-de Sitter universe, the density of the collapsed
halo is $18 \pi^2 \approx 178$ times the critical density of the
universe at collapse time (see also Gunn \& Gott 1972; Bertschinger
1985; Cole \& Lacey 1996), but depends on the density of universe
through the parameter $x = 1 - \Omega(z)$.  Expressing the result in
terms of the super critical density parameter $\Delta_c$
\begin{equation}
\frac{M}{\frac{4}{3}\pi r_{vir}^3} = \Delta_c \rho_c
\end{equation}
Bryan \& Norman (1998) found that for $\Omega + \Omega_{\Lambda} = 1$,
\begin{equation}
\Delta_c \approx 18 \pi^2 + 82 x - 39 x^2 
\end{equation}
for $\Omega_{\Lambda} = 0$,
\begin{equation}
\Delta_c \approx 18 \pi^2 + 60 x - 32 x^2
\end{equation}
where $x = \Omega (z) - 1$.  

Using the virial theorem, it is possible to write equations to relate
the mass, radius, and circular velocity of each halo.  As in
Somerville \& Primack (1998), it can be shown that
\begin{equation}
V_{vir} ^2 = \frac{GM}{r_{vir}} - \frac{\Omega_{\Lambda}}{3} H(z) ^2
r_{vir} ^2
\end{equation}
where $r_{vir}$ is the halo size, $V_{vir}$ is the circular velocity
of the halo at $r_{vir}$, $G$ is Newton's constant, and
\begin{equation}
H(z_f) = H_0 \sqrt{\Omega_{\Lambda,0} + (1 - \Omega_0 - \Omega_{\Lambda,0}) 
	(1 + z_f)^2 + \Omega_0 (1+z_f)^3}.
\end{equation}
Using the fact that $M = \Delta_c \rho_c \frac{4}{3}\pi r_{vir}^3 =
\Delta_c H(z) ^2 \frac{1}{2G} r_{vir}^3$, we can rewrite this as
\begin{equation}
V_{vir} ^2 = \frac{1}{2}(\Delta_c - \Omega_{\Lambda}) H(z) ^2 r_{vir} ^2
\end{equation}
or
\begin{equation}
r_{vir} = \frac{V_{vir}}{\sqrt{\frac{1}{2}(\Delta_c - \Omega_{\Lambda})}H(z)}
\end{equation}
Similarly, we can now rewrite the halo mass as
\begin{equation}
M = \frac{V_{vir} ^2 r_{vir}}{G} = \frac{V_{vir} ^3}
{G H(z) \sqrt{\frac{1}{2}(\Delta_c - \Omega_{\Lambda})}},
\end{equation}
Assuming the matter which settles in the disk to be some fraction
$m_d$ of the halo mass and the angular momentum of this settling
matter to be some fraction $j_d$ of the halo's angular momentum, a
straightforward derivation (e.g., Mo et al.\ 1998a) allows one to
obtain
\begin{equation}
M_d = \frac{m_d V_{vir} ^3}
{G H(z_f) \sqrt{\frac{1}{2}(\Delta_c - \Omega_{\Lambda})}}
\end{equation}
for the mass of the disk and
\begin{equation}
R_d = \frac{1}{\sqrt{2}} \left( \frac{j_d}{m_d} \right ) \lambda r_{vir}
\end{equation}
for the radius of the disk.  The dimensionless angular momentum parameter
$\lambda$ is defined as
\begin{equation}
\lambda = J |E|^{1/2} G^{-1} M^{-5/2}
\end{equation}
where $J$ is angular momentum, $M$ is the mass, and $E$ is the total
energy of the bound system.

There are three relatively simple reasons to go beyond this simple
approach.  First, the adiabatic contraction of the halo due to
dissipation of baryons towards the halo center will modify the halo
profile.  Second, numerical simulations show that model halos actually
have a Navarro, Frenk, \& White (1997) profile for the the dark halo
rather than the isothermal profile used above.  Finally, in order to
make comparisons back to the observations, it is important to consider
the observationally-measured rotational velocities of the disk rather
than the rotational velocities of the halo proper.  Mo et al.\ (1998a)
have found approximate fitting formulas for the consequent corrections
made to the disk radius $R_d$ and the circular velocity at $3R_d$:
\begin{equation}
R_d = \frac{1}{\sqrt{2}} \left( \frac{j_d}{m_d} \right ) \lambda r_{vir}
f_c ^ {-1/2} f_R
\end{equation}
\begin{equation}
V_c (3R_d) = V_{vir} f_V
\end{equation}
where approximate fitting functions for $f_c$, $f_R$, and $f_V$ are
given by
\begin{equation}
f_c \approx \frac{2}{3} + \left( \frac{c}{21.5} \right) ^{0.7}
\end{equation}
\begin{equation}
f_R \approx \left( \frac{\lambda}{0.1} \right) 
^{-0.06+2.71 m_d + 0.0047/\lambda}
(1 - 3 m_d + 5.2 m_d^2) (1 - 0.019c + 0.00025c^2 + 0.52/c)
\end{equation}
\begin{equation}
f_V \approx \left( \frac{\lambda}{0.1} \right) 
^{-2.67 m_d - 0.0038/\lambda + 0.2 \lambda}
(1 + 4.35 m_d - 3.76 m_d^2) \frac{1 + 0.057c - 0.00034 c^2 - 1.54/c}
{\left[ -c/(1+c) + \ln(1+c) \right]^{1/2}}
\end{equation}
where $c$ is the standard halo concentration parameter for the Navarro
et al.\ (1997) profile.  For simplicity, we use $m_d = 0.05$ and
$c=10$ to convert $V_{vir}$ to $V_c$ in order to compare with the
observations.  We do not use Eqs. (15)-(16) for these comparisons.  Note
that larger values of $m_d$ render disks unstable at relatively faint
surface brightnesses and thus have difficulty accounting for Freeman
Law-type surface brightnesses (Freeman 1970).

On the basis of these simple halo scaling relations, the size and mass
of disks at any redshift simply scales as
$1/H(z_f)\sqrt{\frac{1}{2}\Delta_c(z_f)-\Lambda(z_f)}$, $z_f$ being
the redshift at which these high-redshift disks \textit{formed}.
Consequently, the size and luminosity scale as
\begin{equation}
r(z) = \frac{V_{vir}}
{\sqrt{\frac{1}{2}(\Delta_c(z_f(z)) - \Omega_{\Lambda}(z_f(z)))}H(z_f(z))}
\end{equation}
\begin{equation}
L(z) = \frac{L}{M}(z) M(z) = \frac{L}{M}(z) \frac{V_{vir} ^3}
{G H(z_f(z)) \sqrt{\frac{1}{2}(\Delta_c(z_f(z)) - \Omega_{\Lambda}(z_f(z))}} f_V
\end{equation}
Using Eq. (3) and (4), we now have our functions $E_R(z)$ and $E_L(z)$:
\begin{equation}
E_R(z) = \frac{H(z_f(0))\sqrt{\frac{1}{2}(\Delta_c(z_f(0)) -
\Omega_{\Lambda}(z_f(0)))}}{H(z_f(z))
\sqrt{\frac{1}{2}(\Delta_c(z_f(z)) - \Omega_{\Lambda}(z_f(z)))}}
\end{equation}
\begin{equation}
E_L(z) = \frac{\gamma(0) H(z_f(0))\sqrt{\frac{1}{2}(\Delta_c(z_f(0)) -
\Omega_{\Lambda}(z_f(0)))}}{\gamma(z) H(z_f(z))
\sqrt{\frac{1}{2}(\Delta_c(z_f(z)) - \Omega_{\Lambda}(z_f(z)))}}
\end{equation}
where $\gamma(z)$ is the mass-to-light ratio at redshift $z$.  Note
that this is quite different from scaling these disks simply in terms
of the redshift at which these disks were \textit{observed},
particularly in the case of low $\Omega$ where little evolution in the
size or baryonic mass of the disk population is expected.

We assume that the $z=0$ luminosity function scales in number as a
function of $z$ in an analogous way to how the $10^{12} M_{\odot}$
halos scale in number.  Using the Press-Schechter (Press \& Schechter
1974) mass function
\begin{equation}
N(M,z) dM = -\sqrt{\frac{2}{\pi}} \frac{\rho_0}{M}
\frac{\delta_c}{\sigma(M) D(z)} 
\exp \left(
-\frac{\delta_c^2}{2 \sigma^2(M) D(z)^2}
\right) \frac{d\sigma(M)}{dM} dM
\end{equation}
we see that the halo number density scales as
\begin{equation}
n \propto [D(z)]^{-1}
\end{equation}
since the exponential factor remains approximately unity for the
$10^{12} M_{\odot}$ mass scale.  In terms of the formalism of
Eqs. (3-5),
\begin{equation}
E_N(z) = \frac{D(0)}{D(z)}
\end{equation}
Here, $D(z)$, the growth factor, was computed using the formula
tabulated in Carroll, Press, \& Turner (1992).

We now provide a theoretical and observational justification for our
size-luminosity distribution.  Theoretically, in the Fall \& Estathiou
(1980) picture, the spread in surface brightnesses derives from the
spread in dimensionless angular momenta for halos.  An approximate
parameterization of the dimensionless angular momentum distribution is
\begin{equation}
p(\lambda) = \frac{1}{\sqrt{2\pi \sigma_{\lambda}}} \exp 
\left[ - \frac{\ln(\lambda / \bar{\lambda})^2}{2 \sigma_{\lambda} ^2} \right ]
\frac{d\lambda}{\lambda}
\end{equation}
For $\bar{\lambda}=0.05$ and $\sigma_{\lambda}=0.5$, the above
expression closely approximates the distribution obtained from N-body
simulations (Warren et al.\ 1992; Cole \& Lacey 1996; Catelan \&
Theuns 1996) and analytical treatments (Steinmetz \& Bartelmann 1995).
The above spread in dimensionless angular momentum directly translates
into the following distribution of sizes:
\begin{equation}
\phi(r) dr = \frac{1}{\sqrt{2\pi \sigma_{\lambda}}} \exp 
\left[ - \frac{\ln(r / \bar{r_e})^2}{2 \sigma_{\lambda} ^2} \right ]
d \log r
\end{equation}
For disks with a constant mass-to-light ratio, it follows from
Eqs. (13-16) that $r_d \propto r_{vir} \propto V_{vir} H(z_f)^{-1}
\propto M_d^{1/3} H(z_f)^{-2/3} \propto L_d^{1/3} H(z_f)^{-2/3}$.
Ignoring the dependence of $H(z_f)^{-2/3}$ on the luminosity, it
follows that the surface brightness ($L_d / r_d^2$) scales as
$L_d^{1/3}$.

In fact, de Jong \& Lacey (1998) found that the Mathewson, Ford, \&
Buchhorn (1992) data set gave a good fit to the following bivariate
size-luminosity distribution with similar properties to those
predicted above:
\begin{eqnarray}
\lefteqn{
\Phi (r_e,M) d \log r_e dM = \frac{\Phi_0}{\sigma_{\lambda}
\sqrt{2\pi}} 
\exp(-\frac{1}{2} [\frac{\log r_e/r_e ^{*} - 0.4(M-M_*)(2/\beta-1)}
{\sigma_{\lambda}/\ln(10)}]^2)}\nonumber\\
& & 10^{-0.4*(M-M_{*})(\alpha+1)} \exp(-10^{-0.4*(M-M_*)}) d \log r_e dM
\end{eqnarray}
where $\Phi_0 = 0.0033 \textrm{Mpc}^{-3}$, $\alpha = -1.04$, $\beta =
3$, $M_{*} = -22.8$, $r_e ^{*} = 7.9 \textrm{kpc}$, and
$\sigma_{\lambda} = 0.37$ (converting their sizes and luminosities
from $h_0 = 0.65$ to the $h_0 = 0.50$ used here).  Implicit in the
above bivariate distribution is a distribution in sizes analogous to
Eq. (31), a Schechter distribution in luminosity, and a
$SB \propto L^{1/3}$ correlation between luminosity and surface
brightness.  Similar scalings are apparent in the McGaugh \& de Blok
(1997) sample.

Now let us compare a typical $L_{*}$ galaxy in this model with the
observations.  Using our stated assumption, a $L_{*}$ galaxy has a
mass of $1.1 \cdot 10^{12} M_{\odot}$.  A typical formation time
occurs at $z = 0.3$.  Using Eqs. (13-14), the circular velocity and
size of the halo is 132 km/s and 272 kpc (compared to the 140 km/s and
241 kpc predicted assuming a constant $200\rho_c$ for the collapse
density as in Mo et al.\ 1998).  Then, using Eq. (18), the size of the
disk is $\sim6.0$ kpc.

This is smaller than the empirical findings of de Jong \& Lacey (1998)
(7.9 kpc), our own comparisons to local observations (\S4.4) (6.9
kpc), and Freeman's Law, which gives 6.9 kpc.  Supposing this to be
due to a slight cut-off at low values of the dimensionless angular
momentum due to disk instabilities (Efstathiou, Lake \& Negroponte
1982; Dalcanton et al.\ 1997; Mo et al.\ 1998a; van den Bosch 1998),
we scale up the size of a typical $L_{*}$ disk galaxy to 6.9 kpc and
reduce the spread in dimensionless angular momenta to
$\sigma_{\lambda} = 0.37$, as found by de Jong \& Lacey (1999) in the
analysis of the Mathewson et al.\ (1992) sample.

Throughout our analysis, we shall take the $\Omega=0.3$,
$\Omega_{\Lambda}=0.7$ model as our preferred fiducial hierarchical
model because of its better correspondence with the evolution in the
number of small disks observed up to $z\sim1$ (Mao et al.\ 1998).  We
evolve the mass-to-light ratio $\gamma(z)$ as $(1+z)^{-0.5}$ to
reproduce the observed evolution in the Tully-Fisher relationship (see
\S4.2).

\subsection{Infall Model (Backwards Approach)}

Instead of trying to determine how the global structural properties of
disks evolve based on the corresponding properties of their ambient
halos, it is also possible to examine a number of local disk galaxies
in great detail and to use detailed models of their observed
properties (gas profiles, stellar profiles, metallicity profiles,
current SFR profiles, age-metallicity relationships) to determine how
galaxies might have evolved to high redshift.

There is no consideration of how individual halos might evolve
backwards in time in these models, both for simplicity and because of
large uncertainties in the local distribution of dark matter.
Consequently, while for the forwards approach, the entire evolution of
disk properties derives from an evolution of the halo properties, the
infall models considered here completely ignore these effects.
Conversely, while the forwards approach presented here ignores issues
related to the manner in which halo gas is converted into stars, for
the infall model, such issues are important.

Naturally, given that one always adopts the observed local universe in
this approach as known, this approach does not explain in and of
itself why the local disk population is as it is.  Indeed, it cannot
since there is no link to the initial conditions.  In this view, for
the hierarchical approach, we adopted the local universe we did
because it was a natural prediction of the model, and for the infall
approach, we adopted it because it agrees with the observations.

We examine such a model for the evolution of local galaxies based upon
the Prantzos \& Aubert (1995) model for the star formation rates,
metallicites, stars, and gas content for the Milky Way disk.  We
previously presented this infall model elsewhere (Bouwens et al.\
1997; Cay\'on et al.\ 1996), and we shall revisit it here.  This model
ignores radial inflows for simplicity and takes the star formation
rate to be proportional to both the gas surface density ($\Sigma_g$)
and the reciprocal of the radius $r$, which for a flat rotation curve
is proportional to the epicyclic frequency:
\begin{equation}
\frac{d\Sigma_{*}(r,t)}{dt} = \frac{\Sigma_g (r,t)}{\tau_g (r)}
\end{equation}
where $\tau_g (r) = [0.3 (r/r_{\odot})^{-1}\,\textrm{Gyr}]^{-1}$.
Physically, such a star formation rate results if the star formation
rate is proportional to the rate at which molecular clouds collide
(Wang \& Silk 1994) or the periodic compression rate (Wyse \& Silk
1989).

For simplicity, the accretion time scale $\tau_{ff}$ was taken to be
independent of radius since a variation in this time scale is not
strongly constrained by the observations (Prantzos \& Aubert 1995).
The spread in Hubble types was then naturally taken to arise from a
spread in this time scale (Cay\'on et al.\ 1996).  The equation for the
evolution of the gas density is then
\begin{equation}
\frac{d\Sigma_g (r)}{dt} = \frac{\Sigma_g (r,T) + \Sigma_{*} (r,T)}{1
   - e^{-T/\tau_{ff}}} \frac{e^{-t/\tau_{ff}}}{\tau_{ff}} -
   \frac{\Sigma_g (r)}{\tau_g}
\end{equation}
where $T$ is the time from the formation of the disk to the present.
Integrating these equations yields the result
\begin{equation}
\Sigma_g (r,t) = \frac{\Sigma_g (r,T) + \Sigma_{*} (r,T)}{1 -
e^{-T/\tau_{ff}}} \frac{e^{-t/\tau_{ff}} - e^{-t/\tau_g}}{\tau_{ff} -
\tau_g} \tau_g
\end{equation}

Given the fact that this model derived from only one galaxy, it is
difficult to know how to extend its evolutionary predictions to
galaxies with different luminosities and surface brightnesses.  One
possible means of extending this model to galaxies beyond the Milky
Way involves simply scaling the star formation rates by the
differential rotation rate.  The change in the star formation rate
would then be proportional to $V_c/R$ or the typical dynamical time
for the disk.  Since $R$ is roughly proportional to $V_c$, there would
be no large change in the time scales as a function of disk mass or
rotational velocity.  Accordingly, Bouwens et al.\ (1997) simply
elected to scale everything in size to reproduce all luminosities
while conserving surface brightness, which is precisely what we have
done here.

To determine the scaling relations for galaxies using the infall
approach, we performed the calculation for each galaxy on a series of
30 different rings varying logarithmically in size, where the smallest
is a circle with radius 0.2 kpc and the largest is a ring of radius 60
kpc with width 12 kpc as done in Bouwens et al.\ (1997) and Roche et
al.\ (1998).  We calculate the evolution in rest-frame $B$ band
magnitudes by evolving each ring separately to keep track of its gas
mass, stellar composition, and metallicity, and we output its colors
using the Bruzual \& Charlot instantaneous-burst metallicity-dependent
spectral synthesis tables compiled in Leitherer et al.\ (1996).

In the rest-frame $B$ band, we found the following scaling
relationships:
\begin{equation}
E_L (z) = 10^{-0.4(-0.6z)}
\end{equation}
\begin{equation}
E_R (z)  = 1 - 0.27z
\end{equation}
Clearly, without number evolution, we take $E_N(z) = 1$.

For the sake of clarity, we note that the present model differs from
the one presented in Bouwens et al.\ (1997) in terms of both the
luminosity functions used and the bulge-to-total distribution assumed.
Furthermore, in the Bouwens et al.\ (1997) study, the preferred values
of the age $T$ and gas-infall time scale $\tau_{ff}$ used in the
Prantzos \& Aubert (1995) study were scaled to reproduce the number
counts.  No such scaling was attempted in the present model and we
simply use the same $\tau_{ff}$ for all disk types.

This model is similar in spirit to the size-luminosity evolution model
presented by Roche et al.\ (1998) based on the infall models of
Chiappini et al.\ (1997), which models the infall, star formation, and
chemical evolution of both the thin and thick disk components.  For
the purposes of illustration, we shall compare the Chiappini et al.\
(1997) infall model to the one just described in \S4.1, after which we
will restrict our consideration to the redshift scalings given by the
Prantzos \& Aubert prescription.  In this model, the star formation
time scale is equal to
\begin{equation}
\tau = \left\{
\begin{array}{ll}
1\, \textrm{Gyr},&r<2\, \textrm{kpc}\\
(0.875r - 0.75)\,\textrm{Gyr},&r\geq2\, \textrm{kpc}\\
\end{array}
\right.
\end{equation}
The star formation commences at $t_{form} = (16\,\textrm{Gyr} - 0.35
\tau)$.  Note that the time scale for star formation here depends on
the radius to the first power as in our model, the preferred Prantzos
\& Aubert (1995) model, and the recent work by Boissier \& Prantzos
(1999).

\section{Observations}

\subsection{Low-Redshift Samples}

We make comparisons against local samples with information on the
size, luminosity, and circular velocity of local galaxies.  Though in
principle we could have just used the Courteau (1997) sample, we
follow Mao et al.\ (1998) in using a compilation of three
different samples for the comparisons which follow to examine the
three two-dimensional relationships.

\textit{de Jong \& van der Kruit Sample:}

The de Jong \& van der Kruit (1994) sample provides a nice sample for
examining the local size/magnitude relationship.  It is selected from
the Uppsala Catalogue of Galaxies (Nilson 1973, hereinafter UGC), over
only $\sim12.5\%$ of the sky and uses only relatively face-on ($b >
0.625$) galaxies (37.5\% of all orientations).  Following de Jong
(1996), we also take it to be diameter-limited in $R$ to galaxies
larger than $2'$ at 24.7 $R$-band $\textrm{mag/arcsec}^2$.  Whereas de
Jong \& van der Kruit (1994) in their treatment of their sample assume
transparent disks, we correct observed magnitudes to an inclination of
$70\deg$ using the Tully \& Fouqu\'e (1985) inclination corrections.

\textit{Courteau (1997) Sample:}

We use the Courteau (1997) sample to calibrate the local $z=0$
$V_c-\textrm{size}$ relationship.  The Courteau (1997) sample contains
304 Sb-Sc galaxies from the UGC with Zwicky magnitudes $m_B < 14.5$,
$R$-band angular diameters larger than $1'$, and $B$-band major axis
$< 4'$.  We take their $v_{opt} = V_c (3.2 R_d)$ as the circular
velocity and the $25\,r\,\textrm{mag/arcsec}^2$ isophote as the
radius.

\textit{Pierce \& Tully (1988) Sample:}

We use the Pierce \& Tully (1988) sample to calibrate the local $z=0$
$V_c-\textrm{luminosity}$ relationship.  The Pierce \& Tully sample
was taken from galaxies in the area of the Ursa Major cluster and is
complete up to $B_T < 13.3$.  It includes all galaxies which are not
elliptical or S0, not more face on than $30\deg$, and not possessing
confused H I profiles.  Note that in this study and in the Vogt et
al.\ (1996,1997) studies to be discussed, the observed absolute
$B$-band magnitudes were corrected to intrinsic (unextincted) values
using the Tully \& Fouqu\'e (1985) inclination corrections.

\subsection{High-Redshift Samples}

\textit{Simard et al.\ (1999) Sample:}

For the magnitude-radius relationship, we use the data presented by
Simard et al.\ (1999).  This data set contains structural information
for $\sim200$ galaxies to $I<23.5$ from 6 different WFPC2 pointings in
the Groth strip ($\sim 30\,\textrm{arcmin}^2$).  Spectra were obtained
for only a fraction of the faint galaxies, but for galaxies with
spectra, there was nearly 100\% redshift identification.  Following
Simard et al.\ (1999), we can quantify the selection effects of this
sample.  The probability that a galaxy with apparent magnitude
$I_{814}$ and radius $r_d$ would fall in the photometric sample is
$S_{UP} (I_{814},r_d)$.  From Figure 4 of Simard et al.\ (1999), we
have approximated this as
\begin{displaymath}
\left\{
\begin{array}{ll} 1,& I_{814} + 5 \log r_d(\prime\prime) < 21,\\
1 - \frac{2}{3}(I_{814} + 5 \log r_d(\prime\prime)), & 21 < I_{814} +
5 \log r_d(\prime\prime) < 22.5,\\ 0, & 22.5 < I_{814} + 5 \log
r_d(\prime\prime).
\end{array}
\right.
\end{displaymath}
(Actually, this provides a steeper surface brightness cut-off than the
selection function given in Simard et al.\ 1999.)  The probability
that a galaxy with apparent magnitude $I_{814}$ and radius $r_d$ would
be selected from the photometric sample for spectroscopic follow-up is
given by $S_{PS} (I_{814},r_d)$, which we have approximated as
\begin{displaymath}
\left\{
\begin{array}{ll} 1,& I_{814} < 19.3,\\
1 - \frac{0.8}{4}(I_{814} - 19.3),& 19.3<I_{814}<23.5,\\
0,&23.5<I_{814}
\end{array}
\right.
\end{displaymath}
(again by eyeballing Figure 4 of Simard et al.\ 1999).  Putting these
two selection effects together, the probability of selecting a galaxy
with apparent magnitude $I_{814}$ and radius $r_d$ is simply the
product of these quantities, namely, $S_{UP} (I_{814},r_d) S_{PS}
(I_{814},r_d)$.  Given our ignorance about the inclinations used in
the Simard et al.\ (1999) study, we assume an average inclination of
$60\deg$ in transforming the absolute magnitudes to an inclination of
$70\deg$ using the Tully \& Fouqu\'e (1985) law, which yields a
correction of $0.27^m$ here.

\textit{Lilly et al.\ (1998) Sample:}

We also use the data from the LDSS2-CFRS sample with HST WFPC2
follow-up to look at the magnitude-radius relationship.  From Table 3
of Brinchmann et al.\ (1998), the effective area of the CFRS portion
of this sample is $0.01377\,\textrm{deg}^2$ ($49\,\textrm{arcmin}^2$).
The survey is magnitude limited to $17.5 < I < 22.5$, where the
magnitudes are isophotal to $28.0\,I_{AB}\,\textrm{mag/arcsec}^2$.
Though the surface brightness limit is quoted as
$24.5\,I_{AB}\,\textrm{mag/arcsec}^2$, we have used the more
conservative surface brightness limit
$23.5\,I_{AB}\,\textrm{mag/arcsec}^2$.  In principle, then, for any
galaxy detected the isophotal magnitude should be approximately equal
to the total magnitude.  Lilly et al.\ (1998) chose to examine that
subset of galaxies from this data-set which were disk-dominated and
had disk scale lengths $>4$ kpc, a sample we shall henceforth refer to
as the Lilly et al.\ (1998) large disk sample.  Given that the central
surface brightness is not strongly correlated with inclination angle,
Lilly et al.\ (1998) concludes that disks are consistent with being
opaque.  We use the Tully \& Fouqu\'e (1985) to transform the listed
absolute magnitudes to an inclination of $70\deg$.

\textit{Vogt et al.\ (1997) Sample:}

In contrast to high-redshift samples with both magnitude and radial
information, high-redshift samples with circular velocity measurements
are considerably smaller and possess less well-defined selection
criteria.  In fact, the Vogt et al.\ (1996,1997) sample with 16
galaxies is the largest such published sample.  The selection criteria
for this sample is still somewhat qualitative and patchy in nature.
It considers galaxies with an inclination greater than $30\deg$,
detectable line emission, undistorted disk morphology, and an extended
profile.  We assume an $I_{814} < 22.5$ magnitude limit as used in the
Vogt et al.\ (1997) sample.

\section{Results}

\subsection{Basic Scalings}

Before getting into a detailed comparison of the models with the
observations, we begin by illustrating the manner in which the sizes
of $L_{*}$ ($M \sim 1.2 \cdot 10^{12} M_{\odot}$) galaxies typically
evolve as a function of redshift for the different models in Figure 1.
We present this evolution in terms of the rest-frame $B$ half-light
radius, the specification of a band and a measure being necessary to
the intrinsic band and profile dependence of evolution in the infall
model.

The hierarchical models predict more size evolution than expected from
both infall models considered here.  Of course, the $\Omega = 1$ model
possesses more size evolution than the $\Omega=0.3$;
$\Omega_{\Lambda}=0.7$ model, and the $\Omega=0.3$;
$\Omega_{\Lambda}=0.7$ model more size evolution than the $\Omega=0.1$
model because of the steeper dependence of $1/H(z)$ at the typical
\textit{formation} redshifts of disks to $z\sim1$.  The predictions of
both the Prantzos \& Aubert (1995) and Chiappini et al.\ (1997) models
are quite similar, encouragingly enough given that Prantzos \& Aubert
(1995) and Chiappini et al.\ (1997) models represent completely
independent efforts to model the evolution of the Milky Way galaxy.

Taking $R/V_c$ as the measure of size for a given mass halo, we plot
both the Courteau sample at $z=0$ and the higher-redshift data of Vogt
et al.\ (1996,1997) scaled appropriately so that the Courteau data is
centered on unevolved scale size of a galaxy at $z=0$.  At face value,
a comparison of the Vogt data with the Courteau data indicates that
there has been size evolution from $z=0$ to $z=1$ as any of the models
here would predict (Mao et al.\ 1998).  Nevertheless, the lack of a
strong trend in redshift across the Vogt data-set makes one suspicious
that there may be selection effects at work or even systematic errors
in the measurements of parameters which may produce the observed
differences.  In any case, strong conclusions must await the
compilation of a larger high-redshift dataset, where the selection
effects have been more carefully quantified.

We also present model scalings of the number, luminosity, and surface
brightness expected for $L_{*}$ galaxies in Figure 1.  By
construction, the hierarchical models produce more number evolution
than the infall models, which involve no evolution in number.  Except
for the infall model based upon the Chiappini et al.\ (1997)
prescription, our ``Infall'' model produces similar evolution in
luminosity as the hierarchical models to $z\sim1$.  The infall models
also produce less surface brightness evolution than the hierarchical
models.

\subsection{Tully-Fisher Relationship}

To assess hierarchical and infall models, we compare their predicted
Tully-Fisher relationships with both low and high-redshift
observations in the left panels of Figures 2-3.  For the Monte-Carlo
simulations, we use the same selection effects as already specified
for the low (Pierce \& Tully 1988) and high-redshift (Vogt et al.\
1996, 1997) samples.  We have added the Pierce \& Tully (1992) fit to
these plots for comparison.  Naturally, for both our hierarchical and
infall models, we obtained good agreement with the Pierce \& Tully
(1988) sample since we used that sample to adjust the mass-to-light
ratio (we assumed that a $M_{b_J} = -21$ galaxy had a mass $1.2\cdot
10^{12} M_{\odot}$) and its assumed log-normal scatter (one sigma
scatter of 0.3 dex).  At higher redshift, we again obtain basic
agreement with the Vogt et al.\ (1996,1997) sample for both our infall
and hierarchical models.

\subsection{Size-$V_c$ Relationship}

We also compare our model predictions with the low and high-redshift
observations for the size-$V_c$ relationship in the right panels of
Figures 2-3.  Again, we apply the selection criteria given in \S3 to
the low (Courteau 1997) and high-redshift (Vogt et al.\ 1996,1997)
model results.  At low redshift, we obtain basic agreement with the
observations of Courteau (1997) though there seems to be a slight
shift in our models toward larger sizes.  No adjustment of our models
was made to obtain agreement with the Courteau (1997) sample, and so
this comparison can be considered a self-consistency check on our
$z=0$ models.

At high redshift, sizes for both models are consistent, if not a
little larger than the sizes in the Vogt et al.\ (1996,1997) sample.
One of the most surprising thing about a comparison of the models with
the observations is the significant size evolution observed in the
lowest redshift bin ($z<0.6$) relative to the models.  In fact, as
discussed in relation to Figure 1, the low redshift ($z<0.6$) points
seem to have undergone more size evolution than the high redshift
($z>0.6$) points.  As the low redshift points are primarily low
luminosity galaxies and the high redshift high luminosity galaxies,
this could point to some luminosity-dependent evolutionary trend
though the numbers are still too small to make any claims toward this
end.

\subsection{Size-Magnitude Relationship}

As so often in making low-to-high redshift comparisons, freedom in the
choice of the $z=0$ no-evolution model can be very important in
interpreting the high-redshift results.  In particular, due to the
fact that each redshift bin in the Simard et al.\ (1999) sample
contains galaxies of a particular luminosity, significant evolution in
disk surface brightness would appear to be present, simply as a result
of correlations between luminosity and surface brightness at $z=0$.
There are also important surface brightness selection effects in
constructing the Simard et al.\ (1999) sample.  We illustrate the
importance of these considerations in Figure 4 by presenting a
no-evolution model, hereafter referred to as our fiducial no-evolution
model, identical to our $z=0$ hierarchical and infall models except
there is no-evolution in the disk size, number, surface brightness, or
luminosity, i.e., $E_L(z) = E_N(z) = E_R(z) = 1$.  We also present the
surface brightness distributions recovered from a similar no-evolution
model differing only in its use of the $M_{b_J} = -21$ surface
brightness distribution for all luminosities.  We also present the
surface brightness distributions recovered both by including a less
conservative surface brightness selection
\begin{equation}
S_{UP} (I_{814},r_d (\prime\prime)) =
\left\{
\begin{array}{ll} 1,& I_{814} + 5 \log r_d(\prime\prime) < 21,\\
1 - \frac{1}{3}(I_{814} + 5 \log r_d(\prime\prime)), & 21 < I_{814} +
5 \log r_d(\prime\prime) < 24,\\ 0, & 24 < I_{814} + 5 \log
r_d(\prime\prime)
\end{array}
\right.
\end{equation}
(more resembling the one used by Simard et al.\ (1999)) and without
including surface brightness ($S_{UP}$) selection at all.  Notice the
apparent increase in surface brightness for our fiducial model, where
$SB\propto L^{1/3}$, as a function of z relative to our constant
surface brightness model.  It is interesting to note that there is an
absence of galaxies at surface brightnesses close to the threshold for
detection in the Simard et al.\ (1999) sample.

With these caveats in mind, we compare our model $B$-band surface
brightness distributions with both the low and high redshift
observations in Figure 5.  The models seem to be in rough agreement
with the surface brightness distribution of the observations.  Given
that model populations increase in $B$-band surface brightness by
$\sim1.5^m$ to $z\sim1$ (see Figure 1), this suggests a similar
increase in the surface brightness of disks to $z\sim1$.  The models
themselves show no large differences.  As so often, the uncertainties
in the $z=0$ modeling are large enough to preclude detailed
discrimination among models, especially given the limited high
redshift data sets.

In Figures 6-7, we plot the observed luminosity-size distributions at
low (de Jong \& van der Kruit 1994) and high (Simard et al.\ 1999)
redshift and compare them with those obtained for the
$\Omega=0.3/\Omega_{\Lambda}=0.7$ hierarchical and infall models.  We
show a similar comparison of the hierarchical model with the high
redshift Lilly et al.\ (1998) large disk data-set in Figure 8.  We
also plot the cumulative size and luminosity distributions along the
vertical and horizontal axes, respectively, for both the observations
(histogram) and the models (lines).

As in Figures 4-5, the models predict too many low surface brightness
galaxies relative to the observations.  This also results in too many
large model galaxies and too many low luminosity model galaxies
relative to the observed size and luminosity distributions at
intermediate to low luminosities, especially for the Lilly et al.\
(1998) large disk sample even with our more conservative surface
brightness limit.  For large, luminous galaxies, there is no obvious
change in numbers to high redshift.

In Figures 6-7, there are also relatively large differences in
normalization between the observations and models.  This is apparently
the result of large-scale structure (there are small groups/clusters
at $z\sim0.8$ and $z\sim1.0$ in the Groth Strip).  It is therefore
difficult to make extremely quantitative statements about the
evolution in the number density of galaxies with specific surface
brightnesses, sizes, and luminosities across the redshift intervals
surveyed.

Nevertheless, the abundance of high surface brightness disk galaxies
at high redshifts relative to the model predictions is surely
conspicuous and suggests that there has been a significant increase in
the number of high surface brightness disk galaxies to $z\sim1$ as we
have already argued in comparing the model and observed surface
brightness distributions.  Again, shifting the surface brightness
distribution toward these high surface brightnesses (here by the
$\sim1.5^m$ predicted by the models) is an obvious way of
accommodating this increase.  Of course, any argument based on the
normalization of specific galaxy populations is subject to
considerable uncertainties important when such small contiguous areas
are being probed.

To provide a visual comparison between our no-evolution and
evolutionary models, we include a simulation of a patch of the HDF
($I_{F814W}$, $B_{F450W}$, and $V_{F606W}$) in Figure 9 using our
fiducial no-evolution model (panel a), the $\Omega=1$ hierarchical
model (panel b), and the $\Omega=0.3$,$\Omega_{\Lambda} = 0.7$ infall
model (panel c) for comparison with the actual HDF North and South
(panel d).  Clearly, the lack of high surface brightness galaxies is
apparent in the no-evolution model relative to the HDF and even
somewhat in the apparent in the $\Omega=1$ hierarchical model relative
to the HDF.  Of course, our simulations do not include ellipticals or
peculiars, so the actual HDF will include more bright objects than the
simulations.

\section{Discussion}

There is a real question about a lack of low surface brightness
galaxies relative to our predictions, especially as compared to the
no-evolution model predictions.  This conclusion is somewhat dependent
on the assumed correlation between surface brightness and luminosity
as is evident in Figure 4.  This conclusion is also dependent on the
selection biases against low surface brightness galaxies not being
stronger than those considered here.

There is an extensive literature discussing surface brightness
selection biases (Disney 1976; Allen \& Shu 1979) and various attempts
to derive the bivariate luminosity-surface brightness distribution of
galaxies (McGaugh 1996; Dalcanton et al.\ 1997b; Sprayberry et al.\
1997).  Surface brightness has a particularly strong effect on
isophotal magnitude determinations, especially for low surface
brightness galaxies; and this can introduce significant errors in the
magnitude determinations, so the effective volume probed for these
galaxies is significantly smaller than it is for equivalent luminosity
high surface brightness galaxies (McGaugh 1996).

Simard et al.\ (1999) in a detailed quantification of the selection
effects of the DEEP sample do not consider the effect of surface
brightness on the magnitudes and sizes recovered since typical errors
were found to be $0.2^m$ (Simard 1999, private communication).
Despite the relatively small size of this error, it is not entirely
clear to the present authors that the errors would not become quite
significant for the lowest surface brightness galaxies in the sample,
particularly those just marginally detectable given the chosen object
identification and photometric parameters.  Secondly, Simard et al.\
(1999) considers disk galaxies to be optically thin whereas the
observations of Lilly et al.\ (1998) are more consistent with disks
being optically thick.  Highly-inclined optically thin disks would be
much more detectable than face-on or optically-thick disks.  The
upshot is that at many apparent magnitudes and radii, Simard et al.\
(1998) would suppose that at least some highly inclined galaxies would
be detectable and therefore the selection function $S_{UP}$ there
would be non-zero when in reality if disks were optically thick it
would be zero.  For these reasons, we used a slightly more
conservative selection function in surface brightness than that given
in Figure 4 of Simard et al.\ (1999) (see \S3.2).

Another possibility, not considered here, is that low surface
brightness galaxies might form relatively late, meaning that their
mass-to-light ratios remain relatively large until relatively recent
epochs.  Of course, prima facie, this would seem unlikely given the
apparently constant slope in the Tully-Fisher relationship to faint
magnitudes.

In their own analysis of their sample of $\sim200$ galaxies, Simard et
al.\ (1999) concluded that there had been little evolution in the
surface brightness distribution of disk galaxies when all selection
effects had been carefully considered, quite in contrast to our
estimated $\sim1.5^m$ of $B$-band surface brightness evolution.
Little consideration, however, was paid to the evolution in the total
\textit{numbers} of high surface brightness galaxies.  Here, we find
that the number of high surface brightness galaxies dramatically
exceeds that predicted by the evolutionary models considered here, and
we have argued that this provides evidence for an evolution in the
surface brightness distribution of disk galaxies.

Our interpretation seems to be furthermore supported by the lack of
low surface brightness galaxies relative to our models.  For
no-evolution in the disk surface brightness distribution really to be
present as Simard et al.\ (1999) claims, high-redshift intervals
should have similar numbers of low surface brightness galaxies to
those found in local samples, and these galaxies seem to be deficient,
even with respect to our models which show significant evolution in
surface brightness.

While the conclusions of Simard et al.\ (1999) appear to have been
carefully drawn, we would like to suggest that there are significant
uncertainties in their determination of the mean surface brightnesses
in the lowest redshift intervals and therefore the inferred evolution
in surface brightness due to the small size of the low redshift
samples considered.  By applying the selection effects from the
high-redshift bin identically to all redshift intervals, Simard et
al.\ (1999) restricted their analysis to that fraction of disk
galaxies exceeding the high-redshift surface brightness detection
limit.  Applying these selection criteria uniformly to all low
redshift intervals severely pares down the low-redshift samples and
significantly increases the uncertainty of their average surface
brightness measure.  Given the observed range in observed surface
brightness ($\sim 2\, \textrm{mag/arcsec}^2$) and typical numbers ($\sim
5-6$) for the lowest redshift bins, there is a non-negligible
uncertainty in the average surface brightness at low redshift, $\sim
0.6^m$.

Our estimates of $\sim1.5^m$ of $B$-band surface brightness evolution
are somewhat larger than that inferred by most authors.  Roche et al.\
(1998) found $0.9^m$ of surface brightness evolution from $z\sim0.2$
to $z\sim0.9$, Lilly et al.\ (1998) found $0.8^m$ of surface
brightness evolution in their large disk sample, and Schade et al.\
(1995,1996a) inferred $1.2^m$ and $1.5^m$ respectively to $z\sim0.8$.
Despite different differential measures of surface brightness
evolution, most of these samples give similar values for the mean disk
surface brightness near $z\sim1$: $20.79 \pm 0.17$ for the Roche et
al.\ (1998) sample ($0.65<z$), $19.9 \pm 0.2$ for the Simard et al.\
(1999) sample ($0.9<z<1.1$), $20.7 \pm 0.25$ for the Lilly et al.\
(1998) large disk sample ($0.5<z<0.75$), $20.2 \pm 0.25$ for the
Schade et al.\ (1995) sample ($0.5<z$), and $19.8 \pm 0.1$ for the
Schade et al.\ (1996) sample ($0.5<z<1.1$).  Consequently, differences
in the surface brightness evolution inferred derive from differences
in the $z=0$ surface brightness distributions assumed.  We assume a
distribution consistent with the local data of de Jong \& van der
Kruit (1994) and Mathewson et al.\ (1992) as a baseline for measuring
evolution with a surface brightness peaking faintward of Freeman's Law
($\sim 21.7 B\,\textrm{mag/arcsec}^2$; Freeman 1970) while Schade et
al.\ (1995,1996) simply makes reference to Freeman's Law ($\sim
21.65\,b_j\,\textrm{mag/arcsec}^2$).  Simard et al.\ (1999), Lilly et
al.\ (1998), and Roche et al.\ (1998) measure surface brightness
evolution differentially across their samples.  Possible problems here
are surface brightness selection effects and limited low-redshift
samples.

\section{Summary}

In the present paper, we presented models based on two different
approaches for predicting the evolution in disk properties: a
hierarchical forwards approach, where the evolution in disk properties
follows from corresponding changes in halo properties, and a backwards
approach, where the evolution in disk properties follows from an
infall model providing a close fit to numerous observables for the
Milky Way.  We normalized the models to the local $z=0$ observations,
we made high-redshift predictions for the models, and we compared
these predictions with high-redshift observations.

Our findings are as follows:
\begin{itemize}
\item{The hierarchical and infall models predict relatively similar
amounts of evolution in global properties (size, surface brightness,
and luminosity) for disk galaxies to $z\sim1$.  Clearly,
discriminating between the models will require a careful look at
evolution in number (and therefore surveys over a much larger area)
and/or measurements of certain internal properties, like color, star
formation, or metallicity gradients of high redshift disks.}
\item{There is an apparent lack of low surface brightness galaxies in
the high-redshift observations of Simard et al.\ (1999) and Lilly et
al.\ (1998) as compared to model predictions based on local
observations (Mathewson et al.\ 1992; de Jong \& van der Kruit 1994).}
\item{Our model surface brightness distributions produce relatively
good agreement with the observations, suggesting that the $B$-band
surface brightness has evolved by $\sim1.5^m$ from $z=0$ to $z\sim1$
similar to that found in the models.  This finding is supported by the
fact that there is a significantly larger number of high surface
brightness galaxies than in our model predictions, suggesting that
there has been a significant evolution in number, most easily
accommodated by shifting the mean surface brightness of the disk
population to higher surface brightnesses.  This is contrary to the
conclusion reached by Simard et al.\ (1999) based on the same data.}
\end{itemize}

Here the hierarchical and infall models were presented as competing
models to describe the evolution in the properties of disk galaxies.
If the hierarchical structural paradigm is roughly correct as is
generally supposed, the infall model simply provides a modification of
the basic hierarchical scalings to account for the fact that the gas
infall rate or star formation efficiency is not the same at all radii.
In this sort of scenario, if there is an appreciable formation of
structure at low redshift, a consideration of hierarchical scalings is
probably more appropriate and if there is not, a consideration of
scalings following from infall models is probably more appropriate.
Obviously, at the present time, a complete incorporation of radially
dependent gas infall and star formation scenarios into a hierarchical
paradigm is not merited given the lack of high-redshift data needed to
constrain such hybrid models.

We acknowledge helpful discussions with David Schade and Luc Simard.
We are especially grateful to Laura Cay\'on for helping compile some
of the samples used here, for some useful discussions, and for
providing a critical read of this document.  We thank Stephane
Courteau and Nicole Vogt for sending us their data in electronic form.
This research has been supported in part by grants from NASA and the
NSF.  RJB would like to thank the Oxford astrophysics department for
its hospitality while this work was being carried out.

{}

\newpage

\begin{figure}
\epsscale{0.95}
\plotone{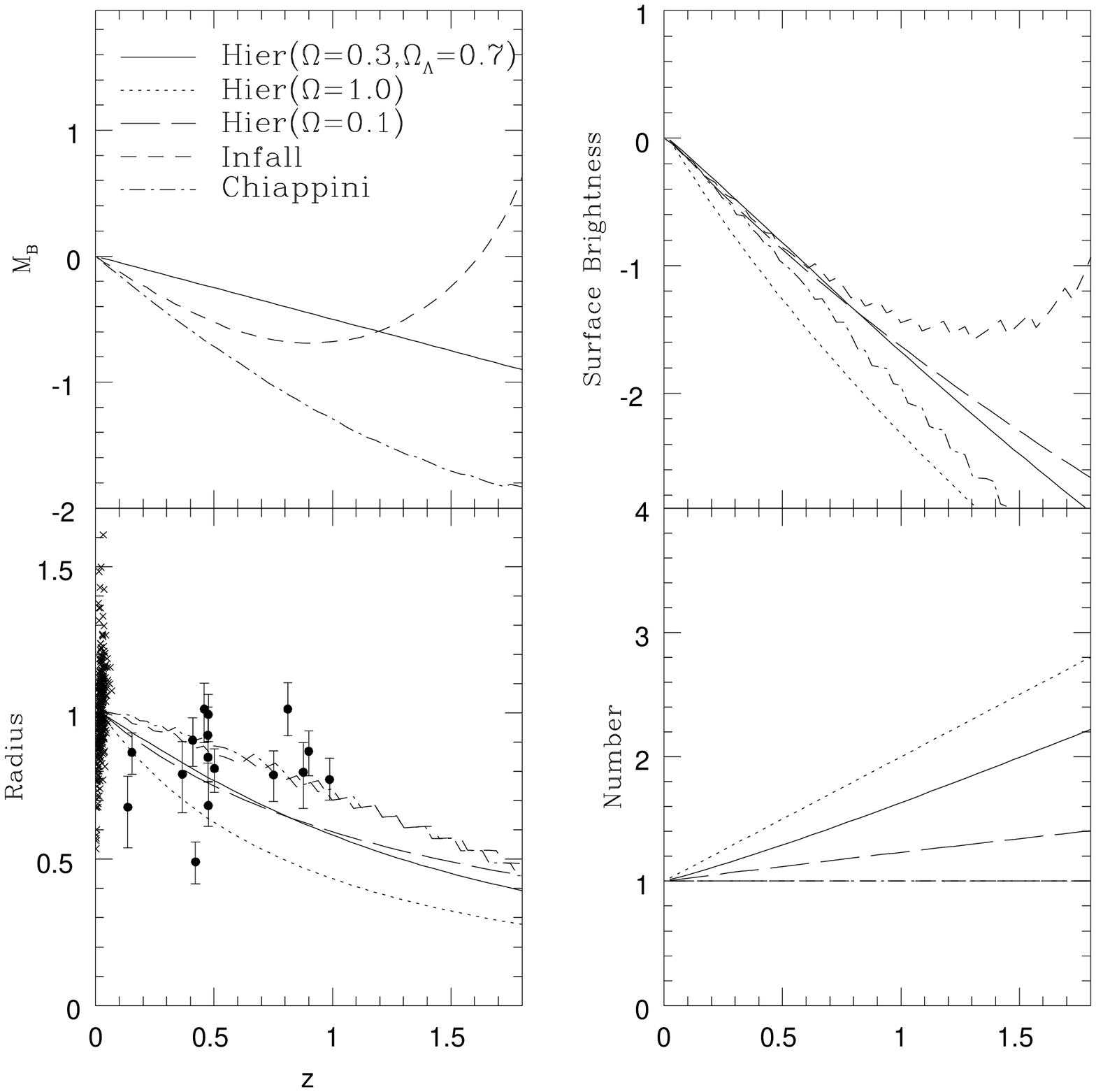}
\caption{Evolution of size, rest-frame $B$ luminosity, number, and
surface brightness for a fiducial $L_{*}$ ($M\sim 1.2 \cdot 10^{12}
M_{\odot}$) galaxy in our $\Omega = 0.3/ \Omega_{\Lambda} = 0.7$
hierarchical model, our $\Omega = 0.1$ hierarchical model (long dashed
line), our $\Omega=1$ hierarchical model (dotted line), our infall
model (dashed line), and the Chiappini et al.\ (1997) infall model
(dotted-dashed line).  Using the $R_d/V_c$ as a measure of size for a
given mass halo, we have added the $z=0$ Courteau (1997) sample and
higher-redshift Vogt et al.\ (1996, 1997) sample to this plot, our
scaling the $R_d/V_c$ values so that the Courteau sample had a
fiducial scale length of unity at $z=0$.  Both the data and models are
presented using $\Omega=0.3$, $\Omega_{\Lambda}=0.7$, $H_0 =
70\,\textrm{km/s/Mpc}$, and corrected to unattenuated magnitudes
(Tully \& Fouqu\'{e} 1985).}
\end{figure}

\newpage

\begin{figure}
\epsscale{0.95}
\plotone{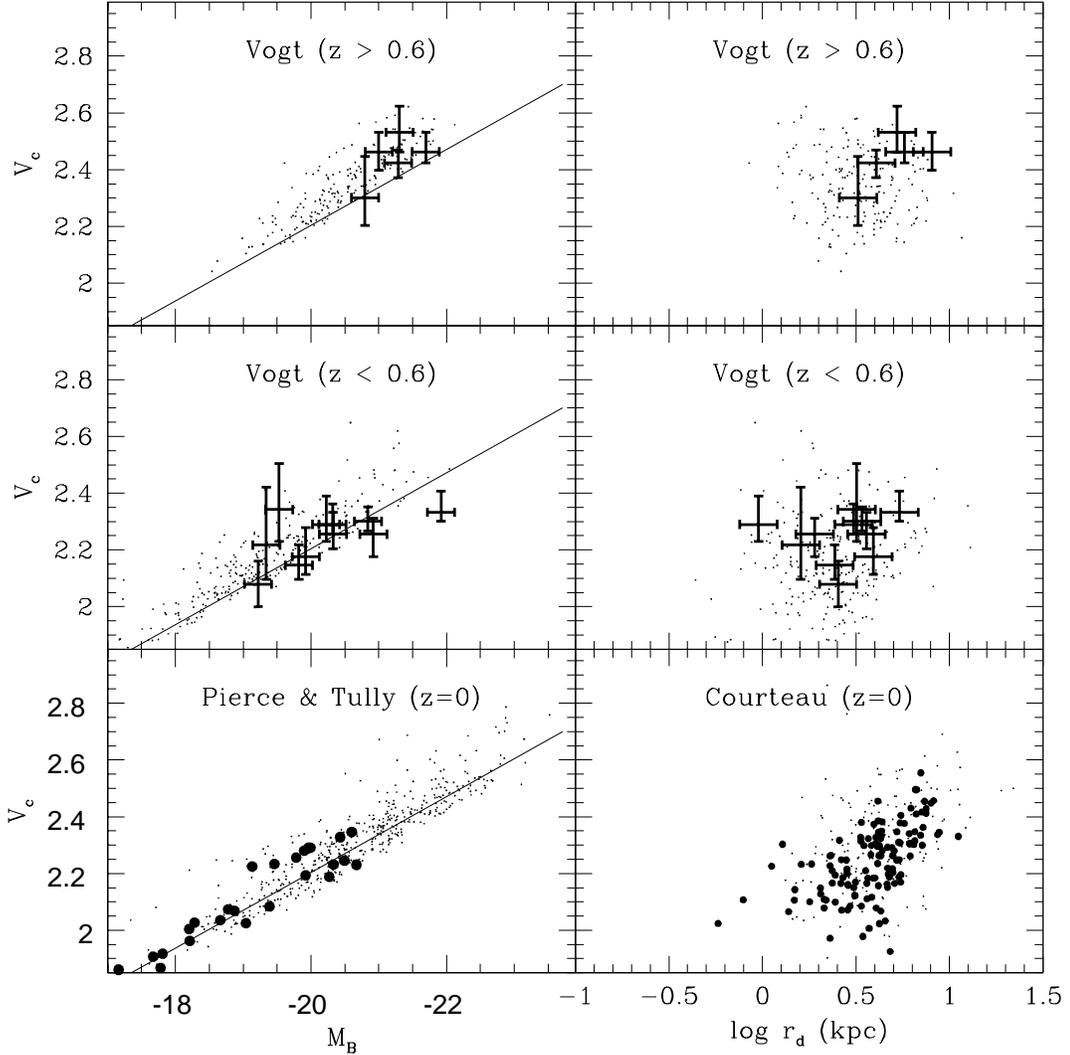}
\caption{Comparison of the observed Tully-Fisher and size-circular
velocity relationships as a function of redshifts against the
presented $\Omega=0.3, \Omega_{\Lambda}=0.7$ hierarchical model at low
and high redshift.  The low-redshift Tully-Fisher data is from Pierce
\& Tully (1988), the high-redshift data is from Vogt et al.\ (1996,
1997), the superimposed line is the Pierce \& Tully (1992)
Tully-Fisher relationship, and the low-redshift $V_c$-size data is
from Courteau (1997).}
\end{figure}

\newpage

\begin{figure}
\epsscale{0.95}
\plotone{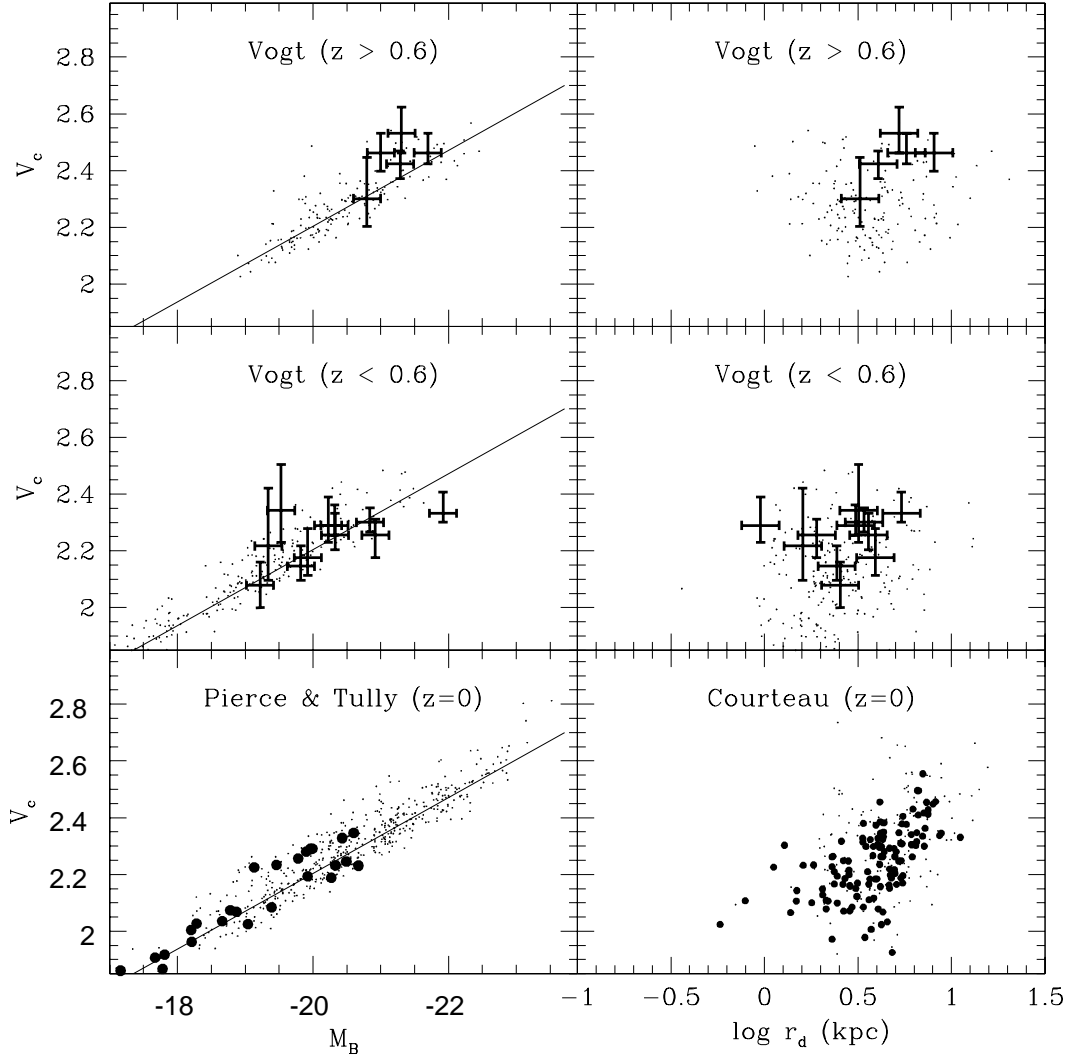}
\caption{Comparison of the observed Tully-Fisher and radius-circular
velocity relationships against the presented $\Omega = 0.3,
\Omega_{\Lambda} = 0.7$ infall model at low and high redshift.  The
data is as in Figure 2.}
\end{figure}

\newpage

\begin{figure}
\epsscale{0.95}
\plotone{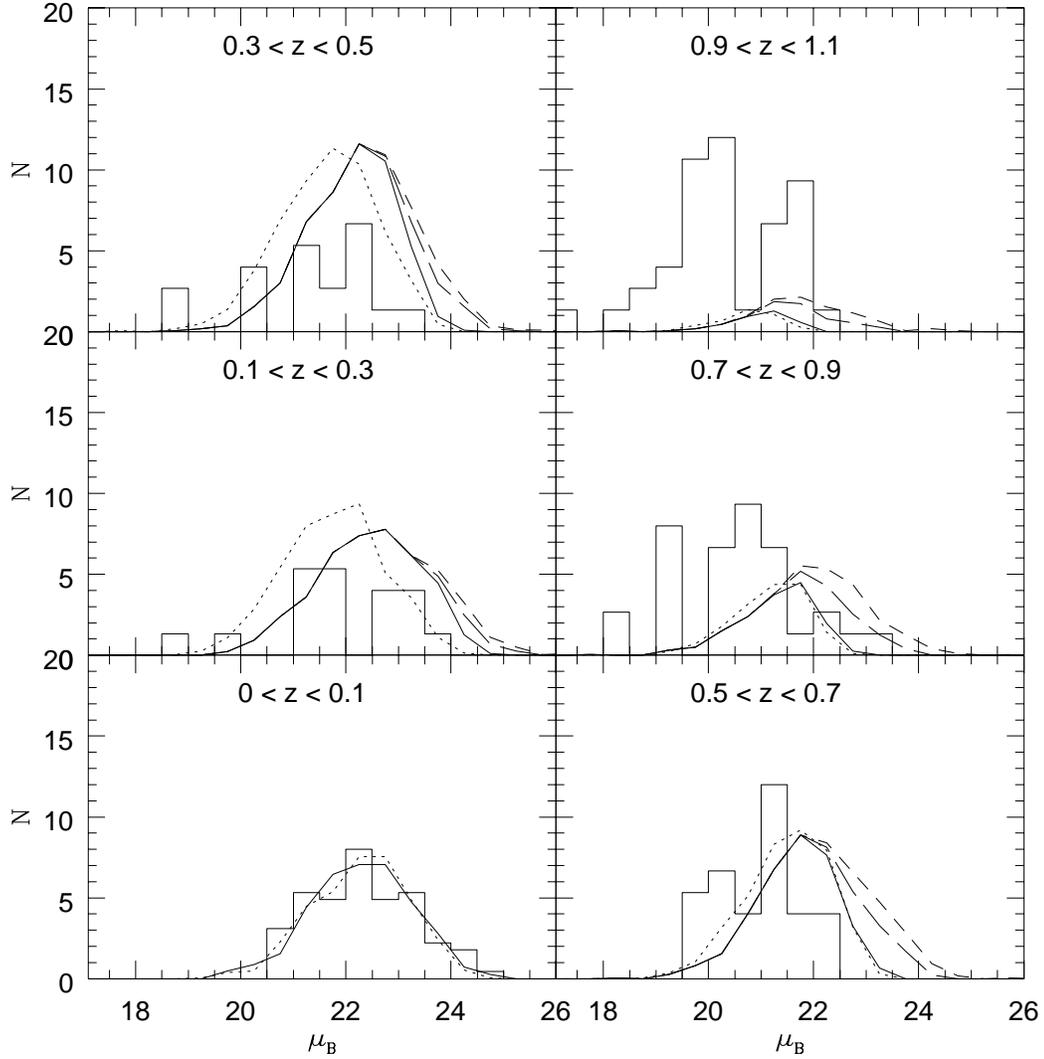}
\caption{Comparison of the observed rest-frame $B$ surface brightness
distributions (histogram) with those from our fiducial no-evolution
model (solid line), our fiducial no-evolution model with constant
$M_{b_J} = -21$ surface brightness distribution (dotted line), our
fiducial no-evolution model without surface brightness selection
(dashed line), and our fiducial no-evolution model with the less
conservative selection function of Simard et al.\ (1999) (long dashed
line).  This illustrates the possible importance of surface brightness
selection and an assumed luminosity-surface brightness correlation on
the conclusions derived.  The local $(z<0.1)$ data is from de Jong \&
van der Kruit (1994) and the high-redshift $(z>0.1)$ data is from
Simard et al.\ (1999).}
\end{figure}

\newpage

\begin{figure}
\epsscale{0.95}
\plotone{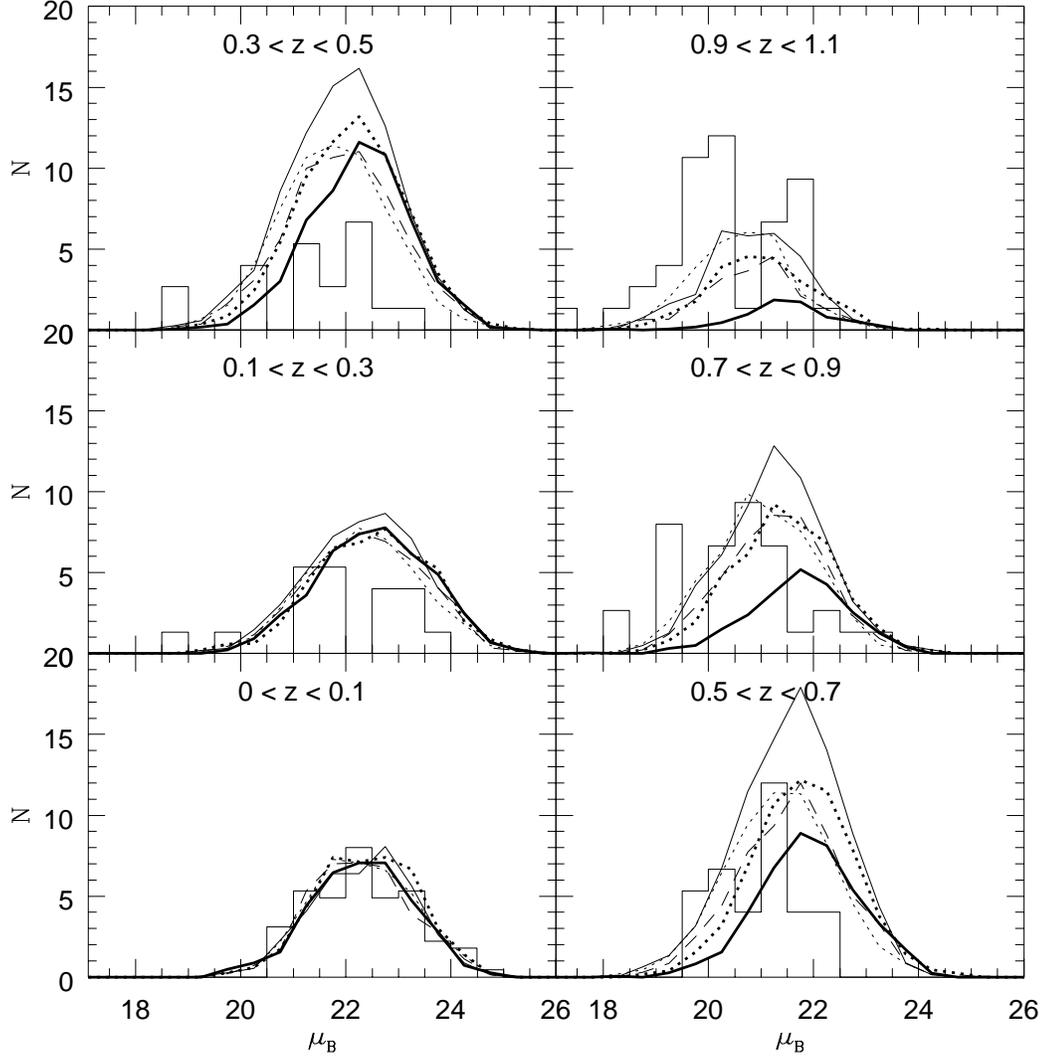}
\caption{Comparison of the observed rest-frame $B$-band surface
brightness distributions (histogram) with those from our hierarchical
models ($\Omega=0.3$, $\Omega_{\Lambda}=0.7$/solid line; $\Omega =
0.1$/dashed line; $\Omega = 1$/dotted line), our infall models (thick
dotted line), and our fiducial no-evolution model (thick solid line)
presented here.  The local $(z<0.1)$ data is from de Jong \& van der
Kruit (1994) and the high-redshift $(z>0.1)$ data is from Simard
et al.\ (1999).}
\end{figure}

\newpage

\begin{figure}
\epsscale{0.95}
\plotone{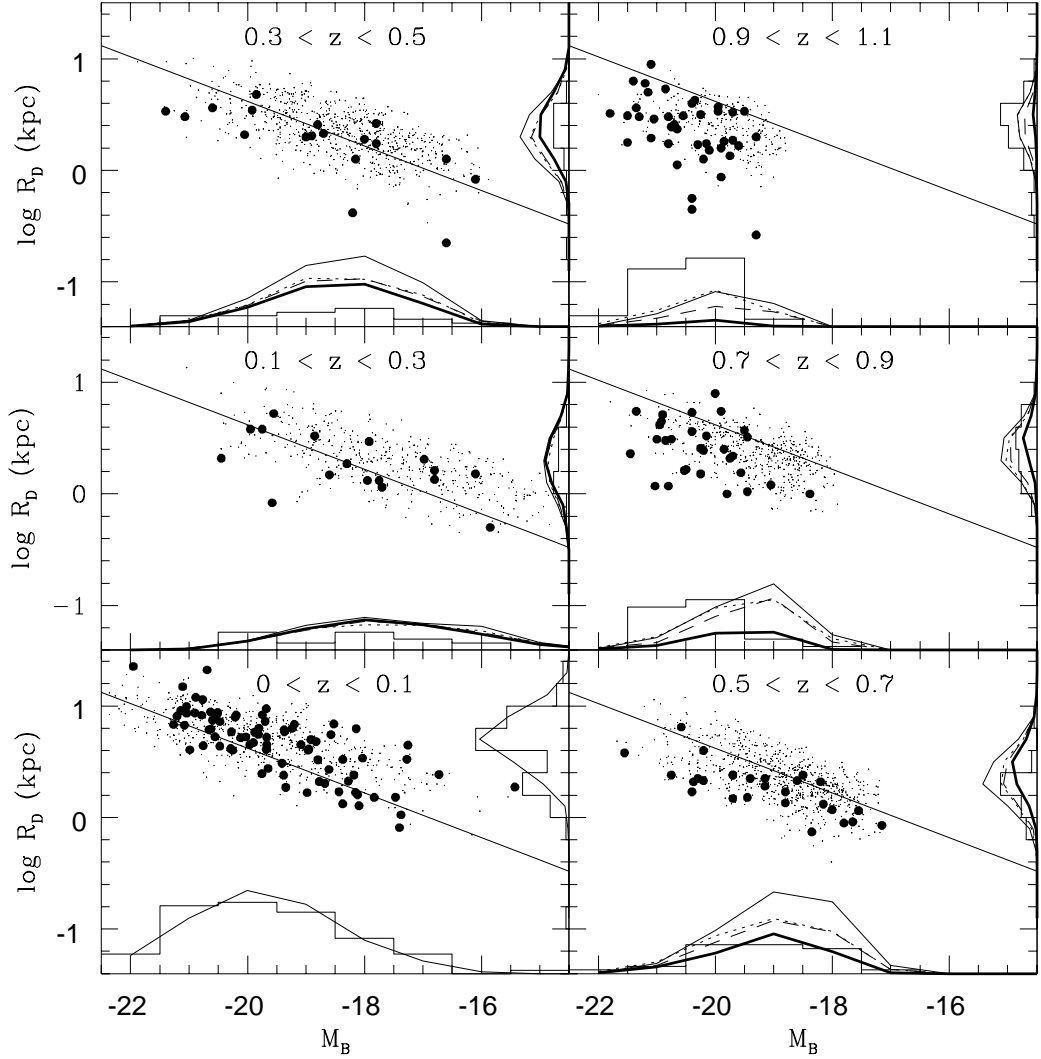}
\caption{Comparison of the observed magnitude-radius (rest-frame $B$)
distributions with the hierarchical models presented here.  The
low-redshift data (filled circles) is from de Jong \& van der Kruit
(1994) and the high-redshift data (filled circles) is from Simard et
al.\ (1999), and the small dots are the results for the $\Omega=0.3$,
$\Omega_{\Lambda}=0.7$ hierarchical model.  Cumulative size and
luminosity distributions are presented on the vertical and horizontal
axes, respectively, for our $\Omega=0.1$ hierarchical model (dashed
line), our $\Omega = 1$ hierarchical model (dotted line), our $\Omega
= 0.3$,$\Omega_{\Lambda}=0.7$ hierarchical model (solid line), and our
fiducial no-evolution model (thick solid line) for comparison with the
observations (histogram).  Both the data and models are presented
using $\Omega=0.3$, $\Omega_{\Lambda}=0.7$, $H_0 =
70\,\textrm{km/s/Mpc}$, and assuming an inclination of 70 deg (Tully
\& Fouqu\'{e} 1985).}
\end{figure}

\newpage

\begin{figure}
\epsscale{0.95}
\plotone{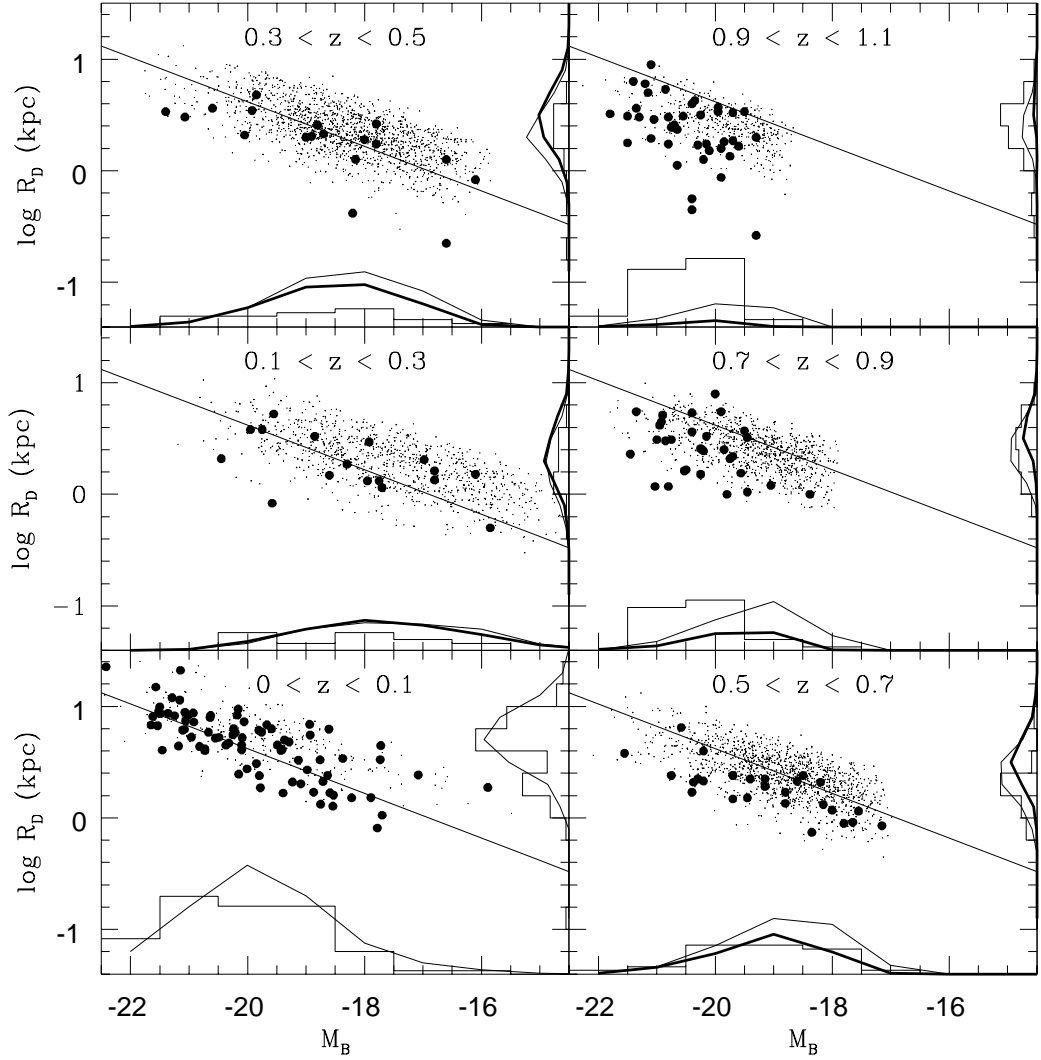}
\caption{Comparison of the observed magnitude-radius (rest-frame $B$)
distributions with the $\Omega = 0.3, \Omega_{\Lambda} = 0.7$ infall
model presented here (\S2.2).  The data is as on Fig 5, and the small
dots trace out the model distribution.  Cumulative size and luminosity
distributions are presented for the infall model (solid lines), the
no-evolution models (thick line), and the observations (histogram).}
\end{figure}

\newpage

\begin{figure}
\epsscale{0.95}
\plotone{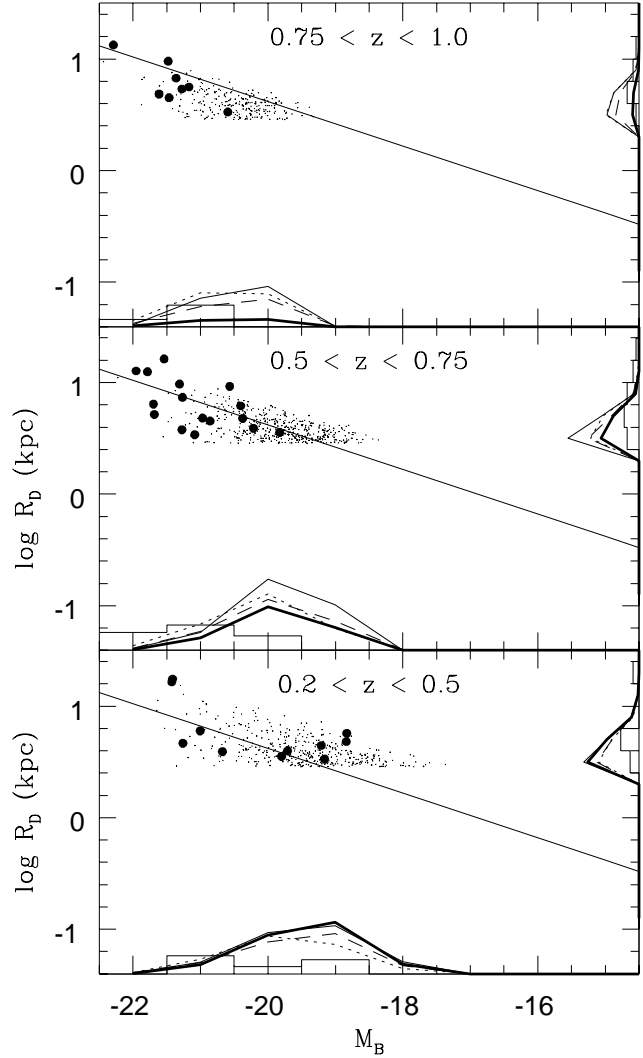}
\caption{Comparison of the observed magnitude-radius (rest-frame $B$)
distributions from the Lilly et al.\ (1998) large disk sample with the
hierarchical models presented here.  The models and data are as in
Figure 6.  Both the data and models are presented using $\Omega=0.3$,
$\Omega_{\Lambda}=0.7$, $H_0 = 70\,\textrm{km/s/Mpc}$, and assuming an
inclination of 70 deg (Tully \& Fouqu\'{e} 1985).  There are an excess
of model galaxies at low magnitudes and large sizes relative to the
observations.}
\end{figure}

\newpage

\begin{figure}
\epsscale{0.8}
\begin{center}
\plotone{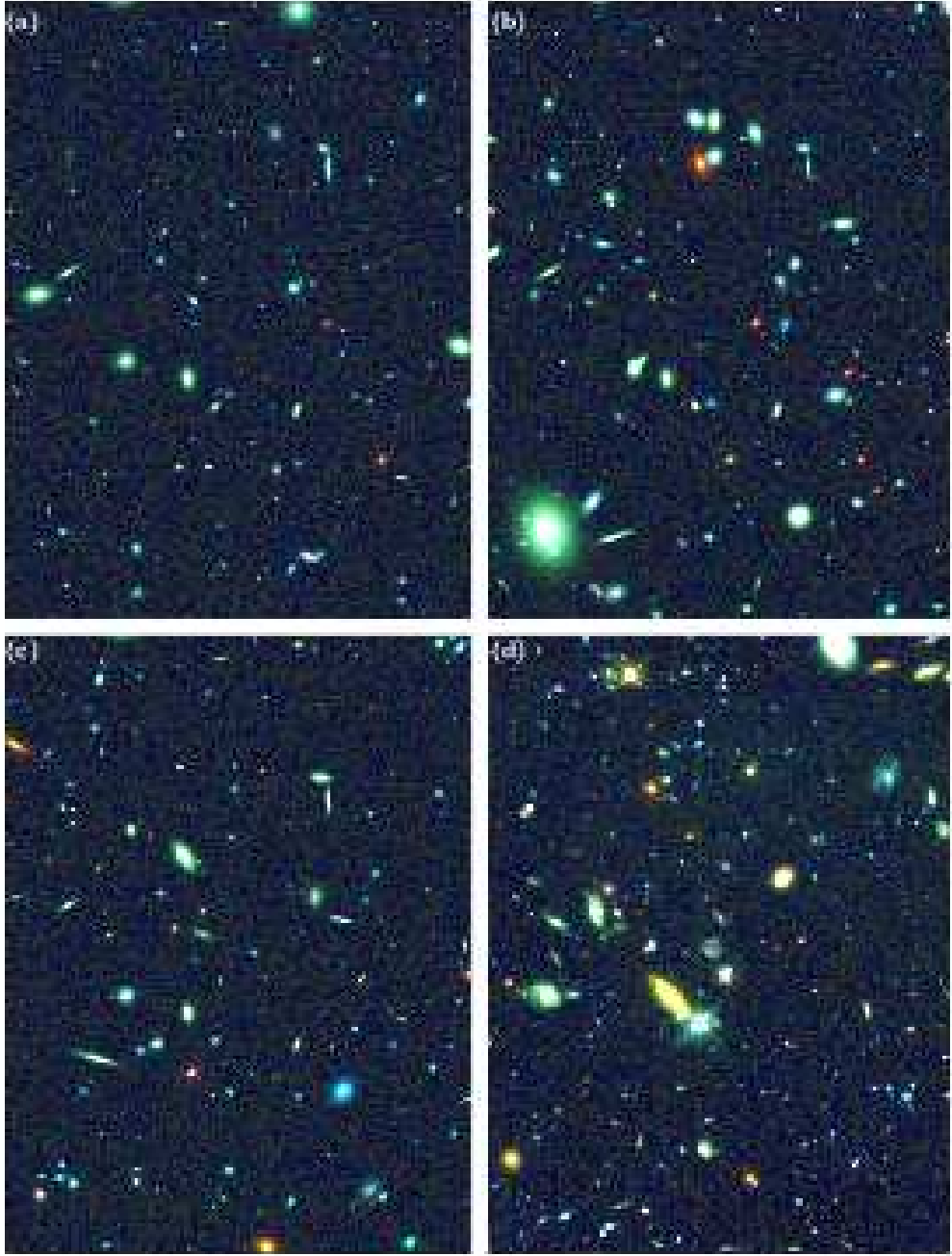}
\end{center}
\caption{Panels (a), (b), (c), and (d) show 60'' x 80'' color images
($I_{F814W}$, $B_{F450W}$, and $V_{F606W}$) for an HDF-depth
simulation using our fiducial no-evolution model, an HDF-depth
simulation using our $\Omega=1$ hierarchical model, an HDF-depth
simulation using our $\Omega=0.3$,$\Omega_{\Lambda}=0.7$ infall model,
and a portion of the actual HDF North and South.  Clearly, our
no-evolution model has fewer high surface brightness galaxies than the
HDF.  Our $\Omega=1$ hierarchical model also appears to lack high
surface brightness galaxies though the fact that we did not include
peculiars and bright ellipticals in our simulations would tend to bias
the eye.}
\end{figure}

\end{document}